\documentclass{mn2e}
\newcommand{\kms}{\ifmmode {\rm km\ s}^{-1} \else km s$^{-1}$\ \fi}
\newcommand{\ergs}{\ifmmode {\rm erg\ s}^{-1} \else erg s$^{-1}$\ \fi}

\newcommand{\feii}{Fe {\sc ii}\ }
\newcommand{\mgii}{Mg {\sc ii}}
\newcommand{\civ}{C {\sc iv}}
\newcommand{\ciii}{C {\sc iii}\ }

\newcommand{\siiv}{Si {\sc iv}\ }

\newcommand{\lb}{\ifmmode L_{\rm Bol} \else $L_{\rm Bol}$\ \fi}
\newcommand{\ledd}{\ifmmode L_{\rm Edd} \else $L_{\rm Edd}$\ \fi}
\newcommand{\leddR}{\ifmmode L_{\rm Bol}/L_{\rm Edd} \else $L_{\rm Bol}/L_{\rm Edd}$\ \fi}
\newcommand{\lx}{\ifmmode L_{\rm 2-10keV} \else  $L_{\rm 2-10keV}$\ \fi}
\newcommand{\hb}{\ifmmode H\beta \else H$\beta$\ \fi}
\newcommand{\ha}{\ifmmode H\alpha \else H$\alpha$\ \fi}
\newcommand{\hg}{\ifmmode H\alpha \else H$\gamma$\ \fi}
\newcommand{\oiii}{[O {\sc iii}]\ }

\newcommand{\mbh}{\ifmmode M_{\rm BH}  \else $M_{\rm BH}$\ \fi}
\newcommand{\lv}{\ifmmode \lambda L_{\lambda}(1350\AA) \else $\lambda L_{\lambda}(1350\AA)$\ \fi}
\newcommand{\lcon}{\ifmmode L_{1350} \else $L_{1350}$\ \fi}

\newcommand{\mdot}{\ifmmode \dot{m} \else \dot{m} \fi }
\newcommand{\llog}{\ifmmode {\rm log} \else {\rm log} \fi }

\usepackage{graphicx}
\usepackage{color}
\usepackage{lscape}

\begin{document}
\title[The Baldwin Effect in QSOs]{The \civ\ Baldwin effect in QSOs from Seventh Data Release of the Sloan Digital Sky Survey }
\author[Bian et al.]
{Wei-Hao Bian$^{1}$, Li-Ling Fang$^{1}$, Ke-Liang Huang$^{1}$, Jian-Min Wang$^{2}$  \\
$^{1}$ Department of Physics and Institute of Theoretical Physics,
Nanjing Normal University, Nanjing
210097, China\\
$^{2}$ Key Laboratory for Particle Astrophysics, Institute of High
Energy Physics, Chinese Academy of Sciences, Beijing 100039, }
\maketitle

\begin{abstract}
Using the properties of SDSS DR7 QSOs catalog from Shen et al., the
Baldwin effect, its slope evolution, the underlying drive for a
large sample of 35019 QSOs with reliable spectral analysis are
investigated. We find that the Baldwin effect exists in this large
QSOs sample, which is almost the same in 11 different redshift bins,
up to $z\sim 5$. The slope is -0.238 by the BCES (\civ\ EW depends
on the continuum), -0.787 by the BCES bisector. For 11
redshift-bins, there is an increasing of the Baldwin effect slope
from $z\sim1.5$ to $z\sim2.0$. From $z\sim2.0$ to $z\sim5.0$, the
slope change is not clear considering their uncertainties or larger
redshift bins. There is a strong correlation between the rest-frame
\civ\ EW and \civ-based \mbh while the relation between the \civ\ EW
and \mgii-based \mbh is very weak. With the correction of \civ-based
\mbh from the \civ\ blueshift relative to \mgii, we suggest that
this strong correlation is due to the bias of the \civ-based \mbh,
with respect to that from the \mgii\ line. Considering the
\mgii-based \mbh, a medium strong correlation is found between the
\civ\ EW and the Eddington ratio, which implies that the Eddington
ratio seems to be a better underlying physical parameter than the
central black hole mass.
\end{abstract}


\section{INTRODUCTION}
Broad emission lines are a prominent property of QSOs. The discovery
of an anti-correlation between the equivalent width (EW) of \civ
$\lambda$1549 emission line and its nearby continuum luminosity in
QSOs rest frame (the Baldwin effect), was first made by Baldwin
(1977) (e.g. see a review by Shields 2006). Over the past 30 years,
a significant amount of effort has been expended to confirm this
effect in \civ\ including other prominent emission lines (e.g.,
Ly$\alpha$, \ciii, \siiv, \mgii, \oiii, Fe K$\alpha$), as well as
exploring its origin and evolution (e.g. Dietrich et al. 2002; Zhou
et al. 2005; Netzer et al. 2006; Kong et al. 2006; Xu et al. 2008;
Wu et al., 2009; Richards et al. 2011).

Although it is believed that the Baldwin effect exists for many
UV/optical emission lines, its origin is still a problem to debate.
Several interpretations about its origin have been proposed (e.g.
Netzer et al. 1992; Dietrich et al. 2002; Shang et al. 2003; Baskin
\& Laor, 2004; Xu et al. 2008; Wu et al., 2009). One promising
interpretation is the softening of the spectral energy distribution
(SED) for increasing luminosity, which lowers the ion populations
having high ionization potentials (e.g., Netzer et al. 1992;
Dietrich et al. 2002).

With the progress in the virial mass of a supermassive black hole
(SMBH, \mbh), it provides the possibility to explore the underlying
physical parameters for the origin of \civ\ Baldwin effect, such as
the Eddington ratio (i.e. the ratio of the bolometric luminosity to
the Eddington luminosity, \leddR), the \mbh, and the luminosity
dependence of metallicity (e.g. Dietrich et al. 2002; Shang et al.
2003; Warner et al., 2004; Baskin \& Laor, 2004, 2005; Xu et al.
2008). The single-epoch virial \mbh can be calculated from the broad
line width (e.g., \hb, \ha, \mgii, \civ) and empirical
luminosity-size relation (e.g. Kaspi et al. 2000; Bian \& Zhao 2004;
Vestergaard \& Peterson 2006; Shen et al. 2011). The \lb can be
calculated from the monochromatic luminosity by the correction
factor estimated from the composite SED (e.g., Richard et al., 2006;
Shen et al. 2011).

For PG QSOs sample, Baskin \& Laor (2004) found a strong correlation
between the \civ\ EW and \leddR, and suggested that the \leddR is the
primary physical parameters driving the Baldwin effect. However,
Shang et al. (2003) used the method of Spectral Principal Component
(SPC) to find there is no correlation between the Baldwin effect and
the \leddR or the \mbh as underlying physical parameters. Nikolajuk
\& Walter (2012) found that weak line QSOs didn't follow the
relation between \civ\ EW and \leddR by Baskin \& Laor (2004).
However, these weak-line QSOs have the same x-ray-optical spectral
index ($\alpha_{ox}$), with respect to other normal QSOs. It was
suggested that the weak-line QSOs are caused by the possible low
covering factor (Nikolajuk \& Walter 2012).

In the past, most of the work was made by using relatively small
samples. The Sloan Digital Sky Survey (SDSS) provides possibility
for us to investigate the Baldwin effect in a larger QSOs sample
(e.g. Xu et al., 2008; Richard et al., 2011). With SDSS DR5, Xu et
al. (2008) used the result from the pipeline of SDSS spectral fit,
which measured the line feature by a single Gaussian fit for the
continuum subtracted spectrum. Xu et al. (2008) found that, up to
$z\approx 5$, the slope of the Baldwin effect seems to have no
effect of cosmological evolution. They also found the \civ\ EW has a
stronger correlation with \civ-based \mbh than the \leddR,
suggesting that the \mbh is probably the primary drive for the
Baldwin effect.

Recently, Shen et al. (2011) made a careful spectral analysis and
gave a compilation of properties for 105783 QSOs from SDSS DR7 QSOs
catalog. Here we use their data to reinvestigate the Baldwin effect
for the \civ\ line and its origin. Our adopted sample is briefly
described in \S 2, the results and the analysis are given in \S 3,
and the conclusions are presented in \S 4.

\section{Sample}

SDSS DR7 covers an imaging area of about 11663 square degrees and a
spectroscopic area of about 9380 square degrees (Abazajian et al.
2009). Schneider et al. (2010) presented a quasars catalog from SDSS
DR7, consisting of 105783 QSOs with $M_{i}<-22$, increasing by 30\%
with respect to that in DR5. The wavelength coverage of SDSS spectrum is
3800\AA\ to 9200\AA. In order to investigate the \civ\ line, we
select 51523 QSOs with $z \geq 1.5$ in SDSS DR7 QSOs.

Shen et al. (2011) carefully considered the UV/optical \feii
contribution, host contribution in low-z QSOs, and multiple-Gaussian
fit for four lines of \ha, \hb, \mgii, \civ\ for the single-epoch
virial SMBH mass calculation. In their SDSS spectral analysis, they
remove the effects of Galactic extinction in the SDSS spectra, and
shift the spectra to rest frame. They compiled these continuum and
line properties, as well as radio, infrared , X-ray properties
(Schneider et al. 2010).

Because adopting the properties of the rest-frame \civ\ EW, \mbh and
\leddR in Shen et al. (2011), we briefly described their spectral
analysis. For the weakness of \feii and not high enough
signal-to-noise (SN) of \civ\ line, Shen et al. (2011) only reported
their \civ\ measurements without the \feii template fits, and
emphasized that their \civ\ EWs may be averagely overestimated by
бл0.05 dex. They fit for the \civ\ line over the $[1500\AA,1600\AA]$
wavelength range, with multiple Gaussians. They didn't subtract the
\civ\ narrow component (e.g., Sulentic et al. 2007). Considering the
possible narrow or broad absorption features of the \civ\ line in
high-z QSOs, they masked out $3 \sigma$ outliers below the 20 pixel
boxcar-smoothed spectrum during the fits for the narrow absorption,
and performed a second fit excluding pixels below $3 \sigma$ of the
first model fit for the broad absorption. They gave the BAL (broad
absorption line) flag for SDSS DR7 QSOs. QSOs flagged as BAL QSOs
are not used in our investigation of the Baldwin effect.

For QSOs with $1.5 \leq z \leq 2.25$, they also fit the \mgii\
emission line, as well as the SMBH mass from \mgii\ line. They fit
for the \mgii\ line over the $[2700\AA,2900\AA]$ wavelength range,
with a single Gaussian (with $\rm FWHM < 1200 \kms $) for the narrow
component, and for the broad component with a single Gaussian or
multiple Gaussians. They provided a new mass calibration from \mgii\
lines by their multiple-Gaussian fits with narrow line subtraction,
which is adopted as the fiducial virial SMBH mass for QSOs with $0.9
\leq z < 1.9$. For $z\geq 1.9$, they adopted the formulae of
Vestergaard \& Peterson (2006) to calculate the \civ\ virial mass.

We use following criteria to select our sample: $z \geq 1.5$,
FWHM(\civ) $>$ 1200 \kms, reduced $\chi^2 < 2.0$, EW(\civ) $>$ 2.0,
SN(\civ) $>$ 5, excluding BAL QSOs. The number of our QSOs sample is
35019 ($\sim 68\%$ of 51523 QSOs). Their distributions of the
redshift, SN, and reduced $\chi^2$ are showed in Figure 1. For the
SMBH mass, we adopted their fiducial viral masses from \mgii\ for
$z<1.9$ and masses from \civ\ for $z>1.9$. The numbers are 14652,
19476, respectively (Shen et al. 2011). In Xu et al. (2008), there
are 26623 QSOs used to investigate the \civ\ Baldwin effect, and
13960 QSOs used to investigate the relation of \civ\ EW with the
accretion parameters (\mbh, \leddR).


\begin{figure}
\begin{center}
\includegraphics[height=8cm,angle=-90]{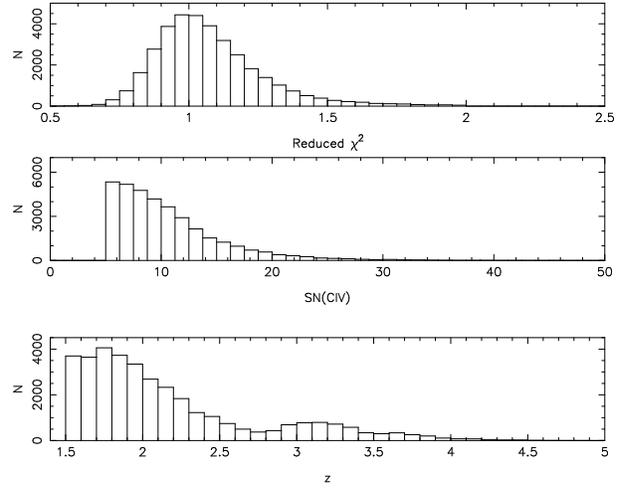} 
\caption{The distributions of the redshift $z$, SN(\civ), and reduced $\chi^2$.}
\end{center}
\end{figure}

\section{RESULTS AND DISCUSSION}
\subsection{The relation between the rest-frame \civ\ EW and \lcon}

\begin{figure}
\begin{center}
\includegraphics[width=6cm,angle=-90]{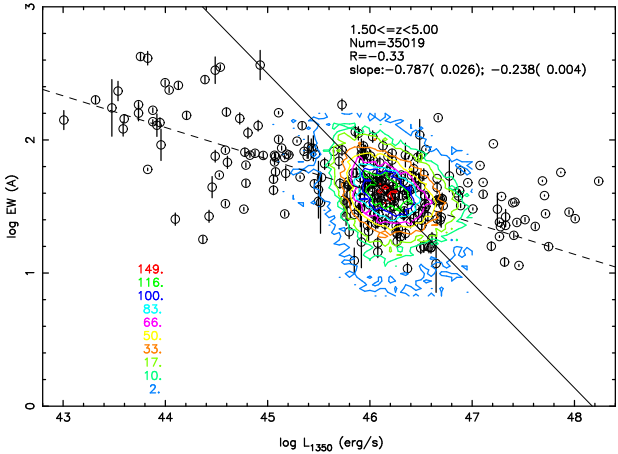} 
\caption{log-log correlation between the \civ\ EW and the continuum
luminosity at 1350\AA\ (in \ergs) for the DR7 sample of 35019 QSOs
with $1.5\leq z < 5.0$. In the left bottom corner, the numbers for
colorful contour levels are shown. The solid line is the linear fit
from the BCES bisector and the bootstrap simulation. The dash line
is BCES with EW(\civ) as the dependent variable and $L_{1350}$ as
the independent variable. In the right top corner, the redshift
coverage, the number of the sample, the spearman's rank correlation
coefficient, and the slopes with BCES($Y|X$) and BCES bisector are
listed. Black circles are data from a compiled sample of Wu et al.
(2009).}
\end{center}
\end{figure}

The following formula is used to express the relation between the
\civ\ EW and the continuum luminosity at 1350 \AA\ (\lcon),
\begin{eqnarray}
\log {\rm EW{\mbox{(\rm \civ)}}}=\alpha+\beta \log \lcon
\end{eqnarray}
Figure 2 gives the relation between the \civ\ EW and \lcon for 35019
QSOs from SDSS DR7. For the total 35019 QSOs, the spearman's rank
correlation coefficient is $R=-0.33$ and the null hypothesis is less
then $10^{-4}$.

The regression analyzes are performed using the bivariate correlated
errors and intrinsic scatter (BCES) algorithm (Akritas \& Bershady
1996). There exists some difference between the BCES($Y|X$) [i.e., Y
= f(X)] and BCES($X|Y$) [i.e., X = g(Y)] regressions. The bisector
bisects these two regressions. The BCES regression is quite robust;
bootstrapping simulations reproduce the theoretically expected
results well (see, e.g., Akritas \& Bershady 1996). For the SDSS
sample, with the BCES($Y|X$) (X is \lcon), the slope is
$\beta=-0.238\pm 0.040$, and the intercept is $\alpha=12.6\pm 0.19$.
However, with the BCES bisector, the slope is $\beta=-0.787\pm
0.026$, and the intercept is $\alpha=37.9\pm 1.20$ (see Table 1).

In Figure 2, we also plot their compiled sample from Wu et al.
(2009). Their compiled sample consists of 272 AGNs: 189 from SDSS,
47 from HST, and 36 from IUE. They used two Gaussian profiles to fit
the \civ\ emission lines after subtracting the power-law continuum.
The continuum luminosity at the rest-frame 2500\AA\ in Wu et al.
(2009) is converted to the luminosity at the rest-frame 1350\AA\ by
assuming $f_{\nu}\propto \nu^{-0.44}$ (i.e, larger by $\sim 0.12
dex$, Vanden Berk et al. 2001). In Figure 2, the data of Wu et al.
(2009) are consistent with our SDSS DR7 sample, as well as the dash
line from BCES($Y|X$). For their compiled sample of Wu et al.
(2009), there are 163 common QSOs in our SDSS DR7 sample. For these
163 common QSOs, the \civ\ EWs from Wu et al. (2009) are consistent
with that from Shen et al. (2009) very well, their \civ\ EW ratio
has a normal distribution with mean $0.01$ and dispersion $0.12$.

For 81 PG QSOs from Baskin \& Laor (2004), 454 Large Bright Quasar
Survey objects by Forster et al. (2001) and 125 pre-COSTAR AGN by
Kuraszkiewicz et al. (2002), we also find that these QSOs/AGNs are
located in the region of SDSS QSOs or consistent with the dash line
from BCES($Y|X$) if they are scaled to our Figure 2.

The Baldwin effect provides a potential method to infer the
luminosity of a QSO from its \civ\ observation. With respect to
classical standard candles of type Ia supernovae (SNe), QSOs can be
observed to a much higher redshift. For our SDSS DR7 sample, we find
the relation between \lcon and rest-frame \civ\ EW by BCES ($Y|X$),
BCES bisector, respectively:
\begin{eqnarray*}
\log \lcon}=(-0.511\pm 0.042)\log \rm EW{\mbox{(\rm \civ)} \\
+ (46.977\pm 0.066)
\end{eqnarray*}

\begin{eqnarray*}
\log \lcon}=(-1.271\pm 0.043)\log \rm EW{\mbox{(\rm \civ)} \\
+ (48.179\pm 0.068)
\end{eqnarray*}
From above formula, we can calculate the rms values of the residuals
after subtracting the predicted luminosity from the measured \civ\
EW. We find that they are 0.286, 0.345, respectively. In order to
use QSOs as standard candles via the Baldwin effect, it is suggested
that we should at least confine the luminosity within an uncertainty
of 30\%, $\sim 0.13$ dex (Wu et al., 2009). Therefore, it is
impossible to use our QSOs data-set as standard candles via the
Baldwin effect.

\begin{table*}
\centering \caption{The \civ\ EW and the continuum luminosity
Regression Parameters. $\log \rm EW{\mbox{(\rm \civ)}}=\alpha+\beta
\log \lcon$.}
\begin{tabular}{lcccccccccc} \hline \hline
$z$  & N  & R  & $\beta^1$ & $\alpha^1$ & $\beta^2$ & $\alpha^2$ & $P_{\rm null}$  \\
(1) & (2) & (3) & (4) & (5) & (6) & (7) & (8) \\
\hline \multicolumn{8}{c} {SDSS DR7 sample}\\
\hline
1.5-5.0  & 35019 & -0.33 & $-0.238\pm 0.040$ &$12.6 \pm 0.19$ & $-0.787\pm 0.026$ &$37.9 \pm 1.20$ &  $1.15\times10^{-25}$ \\
\hline \multicolumn{8}{c} {SDSS DR7 sample in different z bins}\\
\hline
1.5-1.6  & 3699 & -0.33  & $-0.386\pm 0.021$ &$19.4 \pm 0.98$ & $-1.065\pm 0.015$ &$50.2 \pm 0.70$ & $3.90\times10^{-4}$\\
1.6-1.7  & 3645 & -0.32  & $-0.292\pm 0.016$ &$15.0 \pm 0.75$ & $-0.911\pm 0.020$ &$43.5 \pm 0.93$ & $8.93\times10^{-4}$\\
1.7-1.8  & 3727 & -0.35  & $-0.283\pm 0.013$ &$14.6 \pm 0.58$ & $-0.887\pm 0.014$ &$42.4 \pm 0.63$ & $2.65\times10^{-5}$\\
1.8-1.9  & 4057 & -0.35  & $-0.272\pm 0.012$ &$14.1 \pm 0.57$ & $-0.845\pm 0.014$ &$40.5 \pm 0.65$ & $5.29\times10^{-5}$\\
1.9-2.0  & 3343 & -0.31  & $-0.233\pm 0.013$ &$12.3 \pm 0.60$ & $-0.830\pm 0.014$ &$39.9 \pm 0.63$ & $2.73\times10^{-3}$\\
2.0-2.1  & 2687 & -0.32  & $-0.230\pm 0.015$ &$12.2 \pm 0.70$ & $-0.858\pm 0.015$ &$41.2 \pm 0.67$ & $6.10\times10^{-3}$\\
2.1-2.25 & 3265 & -0.33  & $-0.249\pm 0.013$ &$13.1 \pm 0.62$ & $-0.850\pm 0.016$ &$40.9 \pm 0.74$ & $1.32\times10^{-3}$\\
2.25-2.5 & 3173 & -0.34  & $-0.273\pm 0.014$ &$14.2 \pm 0.64$ & $-0.840\pm 0.015$ &$40.4 \pm 0.67$ & $4.99\times10^{-4}$\\
2.5-3.0  & 2740 & -0.34  & $-0.279\pm 0.015$ &$14.5 \pm 0.69$ & $-0.845\pm 0.023$ &$40.7 \pm 1.06$ & $9.22\times10^{-4}$\\
3.0-3.5  & 3213 & -0.30  & $-0.238\pm 0.014$ &$12.6 \pm 0.66$ & $-0.698\pm 0.242$ &$33.9 \pm 11.2$ & $7.52\times10^{-3}$\\
3.5-5.0  & 1470 & -0.27  & $-0.215\pm 0.022$ &$11.5 \pm 1.04$ & $-0.031\pm 0.696$ &$2.93 \pm 32.4$ & $1.59\times10^{-1}$\\
\hline \multicolumn{8}{c} {SDSS DR7 sample in different z bins
($10^{45.6}\le \lcon \le 10^{46.6} \ergs$)}\\
\hline
1.5-1.6  & 3566 & -0.30  & $-0.387\pm 0.023$ &$19.4 \pm 1.05$ & $-1.140\pm 0.015$ &$53.8 \pm 0.70$ & $4.74\times10^{-3}$\\
1.6-1.7  & 3474 & -0.28  & $-0.293\pm 0.018$ &$14.9 \pm 0.83$ & $-0.997\pm 0.018$ &$47.5 \pm 0.81$ & $1.43\times10^{-2}$\\
1.7-1.8  & 3836 & -0.31 & $-0.287\pm 0.016$ &$14.7 \pm 0.74$ & $-0.919\pm 0.012$ &$44.1 \pm 0.55$ & $1.41\times10^{-3}$\\
1.8-1.9  & 3518 & -0.31  & $-0.280\pm 0.016$ &$14.5 \pm 0.73$ & $-0.882\pm 0.052$ &$37.6 \pm 0.24$ & $3.52\times10^{-3}$\\
1.9-2.0  & 3136 & -0.27  & $-0.230\pm 0.016$ &$12.2 \pm 0.75$ & $-0.828\pm 0.011$ &$38.9 \pm 0.53$ & $3.48\times10^{-2}$\\
\hline

\end{tabular}\\
$^1$:BCES($Y|X$) where Y is EW(\civ) as the dependent variable, and
X is $L_{1350}$ as the independent variable. $^2$:BCES bisector
result.
\end{table*}

\subsection{The slope of the Baldwin effect}
In previous studies, the slope of the Baldwin effect is $-0.14 \sim
-0.20$ (see Table 1 in Xu et al. 2008 and reference therein).
Dietrich et al. (2002) found the slope becomes steep for luminous
QSOs , from $-0.14$ for their total sample to  $-0.20$ for QSOs with
$\log \lambda L_{\lambda}(1450 \AA) \ge 44~ \ergs$. We get a slope
from the BCES($Y|X$), $\beta=-0.238\pm 0.04$ (Table 1), which is
steeper than some previous studies. By BCES bisector, the slope is
$\beta=-0.787\pm0.026$. The slope of BCES bisector ($-0.787$) is
more consistent with the slope of the "intrinsic" Baldwin effect,
($\beta=-0.72$) by Pogge \& Peterson (1992). Considering the
compiled sample from Wu et al. (2009), the slope from the
BCES($Y|X$) is better than the slope of BCES bisector when wide
luminosity range for QSOs are concerned (Figure 2).

The redshift coverage of our sample is from 1.5 to 5.0 (see Figure
1). In order to investigate the relation of the Baldwin effect with
the redshift, We divide our sample into 11 bins according to the
redshift, and the redshift bins are 1.5-1.6, 1.6-1.7, 1.7-1.8,
1.8-1.9, 1.9-2.0, 2.0-2.1, 2.1-2.25, 2.25-2.5, 2.5-3.0, 3.0-3.5,
3.5-5.0 (Table 1). The number in different redshift bins is about
3000. For different redshift bins, we calculate the spearman's rank
correlation coefficient, the null possibility, the slopes and the
intercepts with the BCES ($Y|X$) and the BCES bisector (Table 1).
Except the highest bin ($3.5\leq z < 5.0$), We find a medium strong
correlation for different redshift bins ($|R| \geq 0.3$).

It seems that there is an increase of the slope from $z\sim1.5$ to
$z\sim2$ (Figure 3), which is the same to that by the BCES ($Y|X$)
and BCES bisector. From $z\sim2$ to $z\sim5$, the slope evolution is
not clear due to the large uncertainties. If we believe that the
Baldwin effect is due to the softening of SED for increasing
luminosity, the slope evolution from $z\sim 1.5$ to $z\sim 2$
implies the evolution of QSOs SED below $z\sim 2$. It is also
possibly related to the covering factor evolution. In low redshift
bins, the dynamical range in luminosity is larger due to the
flux-limited nature of the sample, which might affect the linear
regression. For redshift bins below $z=2.0$, in the same luminosity
range, i.e., $10^{45.6}$ to $10^{46.6}$ \ergs, we find R is $\sim
-0.3$, and the slopes are consistent with the results for the total
sample in different redshift bins (Table 1).

\begin{figure*}
\begin{center}
\includegraphics[width=6cm,angle=-90]{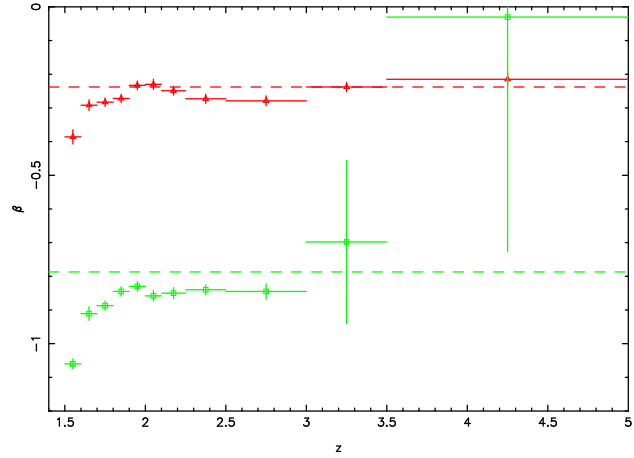} 
\caption{The slope versus the redshift. The red points are for the
slopes by the BCES($Y|X$). The green ones are for the slopes by the
BCES bisector. The red/green dash lines are for the slopes for the
total sample, respectively by the BCES($Y|X$) and bisector.}
\end{center}
\end{figure*}

With SDSS DR5, Xu et al. (2008) used the SpecLine table to
investigate \civ\ Baldwin effect. The SpecLine table is from the
pipeline of SDSS spectral fit. The \civ\ line feature is measured by
a single Gaussian fit for the continuum-subtracted spectrum. Xu et
al. (2008) investigated the \civ\ Baldwin effect in the observed
frame. In Xu et al. (2008), there was no correction of Galactic
extinction for the \lcon and the \civ\ EW was given in the observed
frame. In Figure 4, we showed a comparison of \civ\ EW in the rest
frame for 10232 common QSOs from the current sample and Xu et al.
(2008). We find that, the rest-frame \civ\ EW from the current
sample is averagely larger by 0.23 dex than that from Xu et al.
(2008). The difference is due to the spectral fitting, such as a
single Gaussian fit for \civ\ in Xu et al. (2008), multiple-Gaussian
fit in Shen et al. (2011).

\begin{figure*}
\begin{center}
\includegraphics[width=6cm,angle=-90]{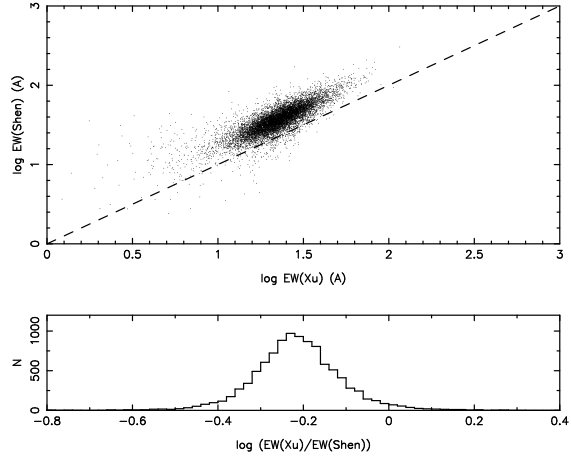} 
\caption{Top: A comparison of the rest-frame \civ\ EW for 10232
common QSOs from the current sample and Xu et al. (2008). The dash
line is 1:1. Bottom: the distribution of the EW ratio of QSOs from
Xu et al. (2008) to the current sample.}
\end{center}
\end{figure*}

\subsection{The relation between \civ\ EW and \mbh or \leddR}

\begin{table*}
\centering \caption{The \civ\ EW drive Regression Parameters. $\log
\rm EW{\mbox{(\rm \civ)}}=\alpha+\beta \log \lcon$, $\log \rm
EW{\mbox{(\rm \civ)}}=\alpha+\beta \log \mbh$, or $\log \rm
EW{\mbox{(\rm \civ)}}=\alpha+\beta \log \leddR$.}
\begin{tabular}{lcccccccccc}
\hline \hline
Drive & $z$  & N  & R  & $\beta^1$ & $\alpha^1$ & $\beta^2$ & $\alpha^2$  \\
(1) & (2) & (3) & (4) & (5) & (6) & (7) & (8) \\
\hline
\lcon & 1.5-5.0  & 35019 & -0.33 & $-0.238\pm 0.040$ &$12.6 \pm 0.19$ & $-0.787\pm 0.026$ &$37.9 \pm 1.20$\\
      & 1.5-1.9  & 15128 & -0.34  & $-0.299\pm 0.076$ &$15.4 \pm 0.35$ & $-0.920 \pm 0.007$ &$44.0 \pm 0.34$\\
      & 1.9-5.0  & 19891 & -0.31  & $-0.226\pm 0.051$ &$12.0 \pm 0.23$ & $-0.755\pm 0.057$ &$36.5 \pm 2.63$\\
\hline
\mbh  &1.5-5.0  & 34128 & -0.27 & $-0.125\pm 0.011$ &$2.74 \pm 0.10$ & $-0.731\pm 0.025$ &$8.34 \pm 0.23$\\
      &1.5-1.9  & 14652 & 0.07  & $0.665\pm 0.087$ &$0.985 \pm 0.08$ & $0.959 \pm 0.057$ &$-7.22 \pm 0.05$\\
      &1.9-5.0  & 19476 & -0.48  & $-0.178\pm 0.019$ &$32.3 \pm 0.18$ & $-0.521\pm 0.052$ &$6.41 \pm 0.49$\\
\hline
\leddR &1.5-5.0  & 34128 & 0.01 & $0.272\pm 0.001$ &$1.58 \pm 0.01$ & $0.182\pm 0.091$ &$1.77 \pm 0.83$    \\
      &1.5-1.9  & 14652 & -0.28  & $-0.268\pm 0.080$ &$1.41 \pm 0.06$ & $-0.858 \pm 0.080$ &$0.992 \pm 0.07$ \\
      &1.9-5.0  & 19476 & 0.26   & $0.0003\pm 0.00009$ &$1.57 \pm 0.01$ & $0.090\pm 0.067$ &$1.69 \pm 0.07$ \\

\hline
\end{tabular}\\
$^1$:BCES($Y|X$) where Y is EW(\civ) as the dependent variable, and X
is \lcon, \mbh, \leddR as the independent variable, respectively.
$^2$:BCES bisector result.
\end{table*}

\begin{figure*}
\begin{center}
\includegraphics[width=6cm,angle=-90]{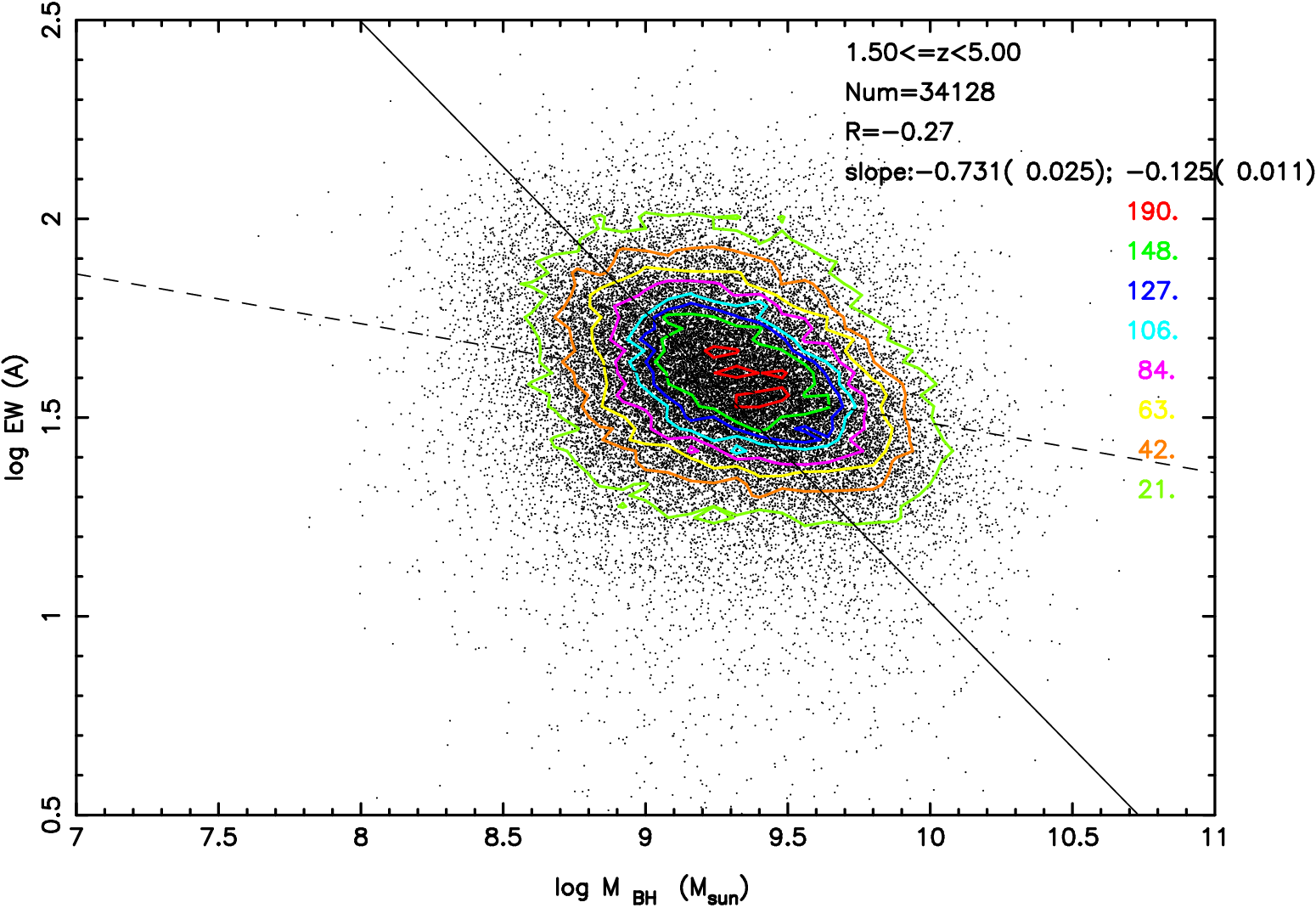} 
\includegraphics[width=6cm,angle=-90]{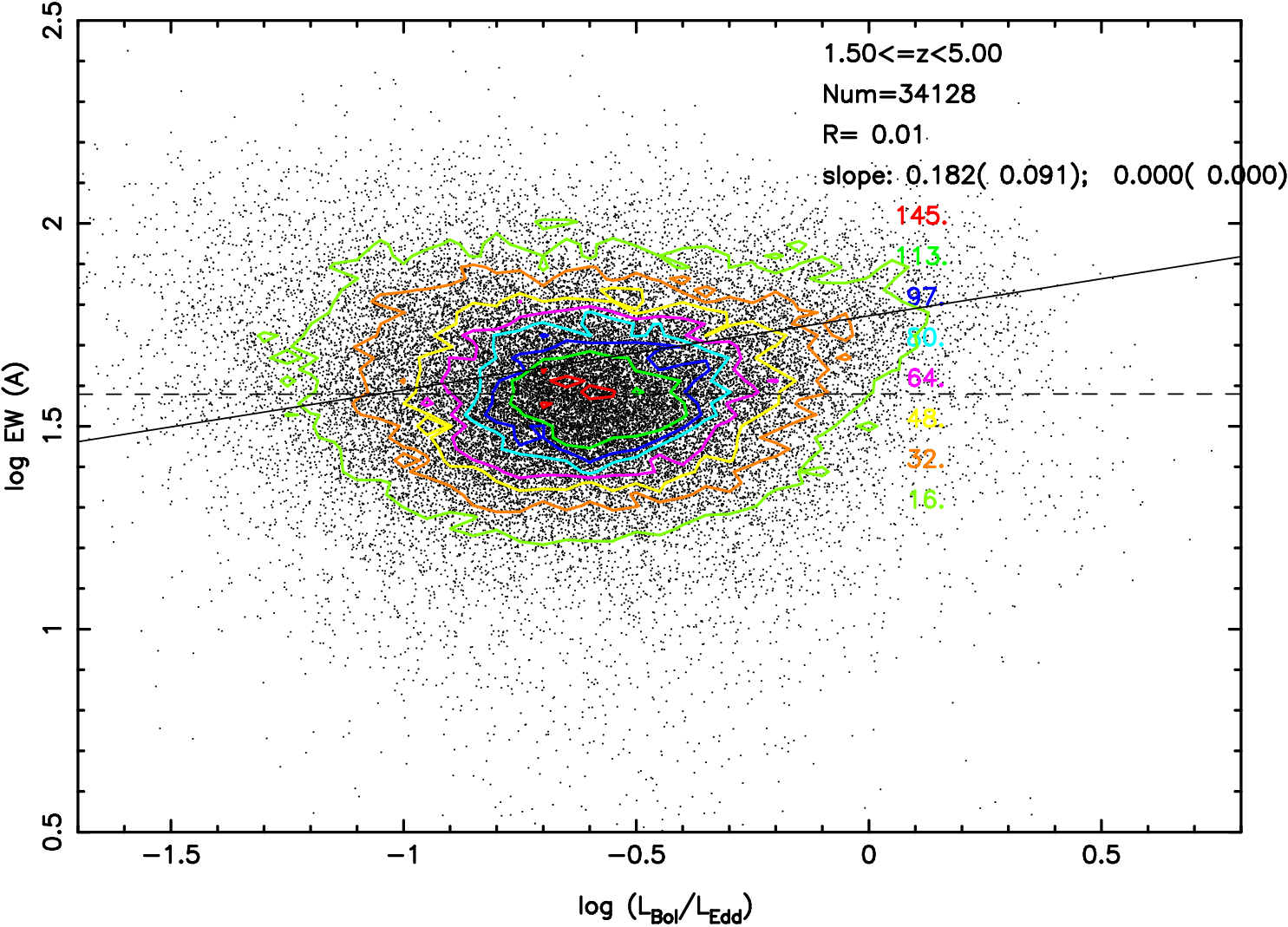} 
\caption{log-log correlation between the rest-frame \civ\ EW and the
\mbh (left), \leddR (right). In the right of two panels, the numbers
for colorful contour levels are shown. The solid line is the linear
fit from the BCES bisector and the bootstrap simulation. The dash
line is BCES with EW(\civ) as the dependent variable and \mbh/\leddR
as the independent variable. In the right top corner, the redshift
coverage, the number of the sample, the spearman's rank correlation
coefficient, the slopes and their errors with BCES($Y|X$) and BCES
bisector are listed.}
\end{center}
\end{figure*}

\begin{figure*}
\begin{center}
\includegraphics[width=6cm,angle=-90]{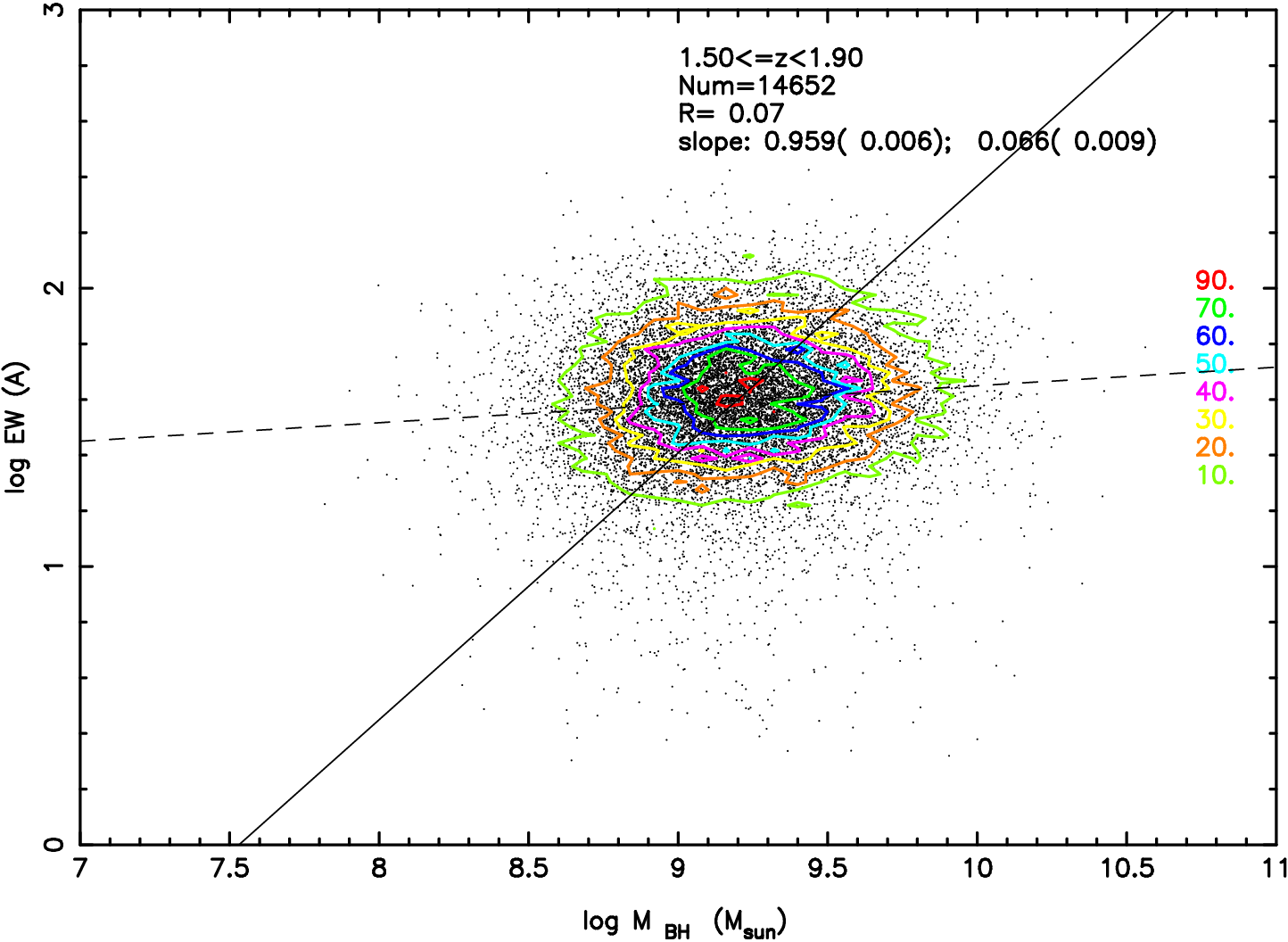} 
\includegraphics[width=6cm,angle=-90]{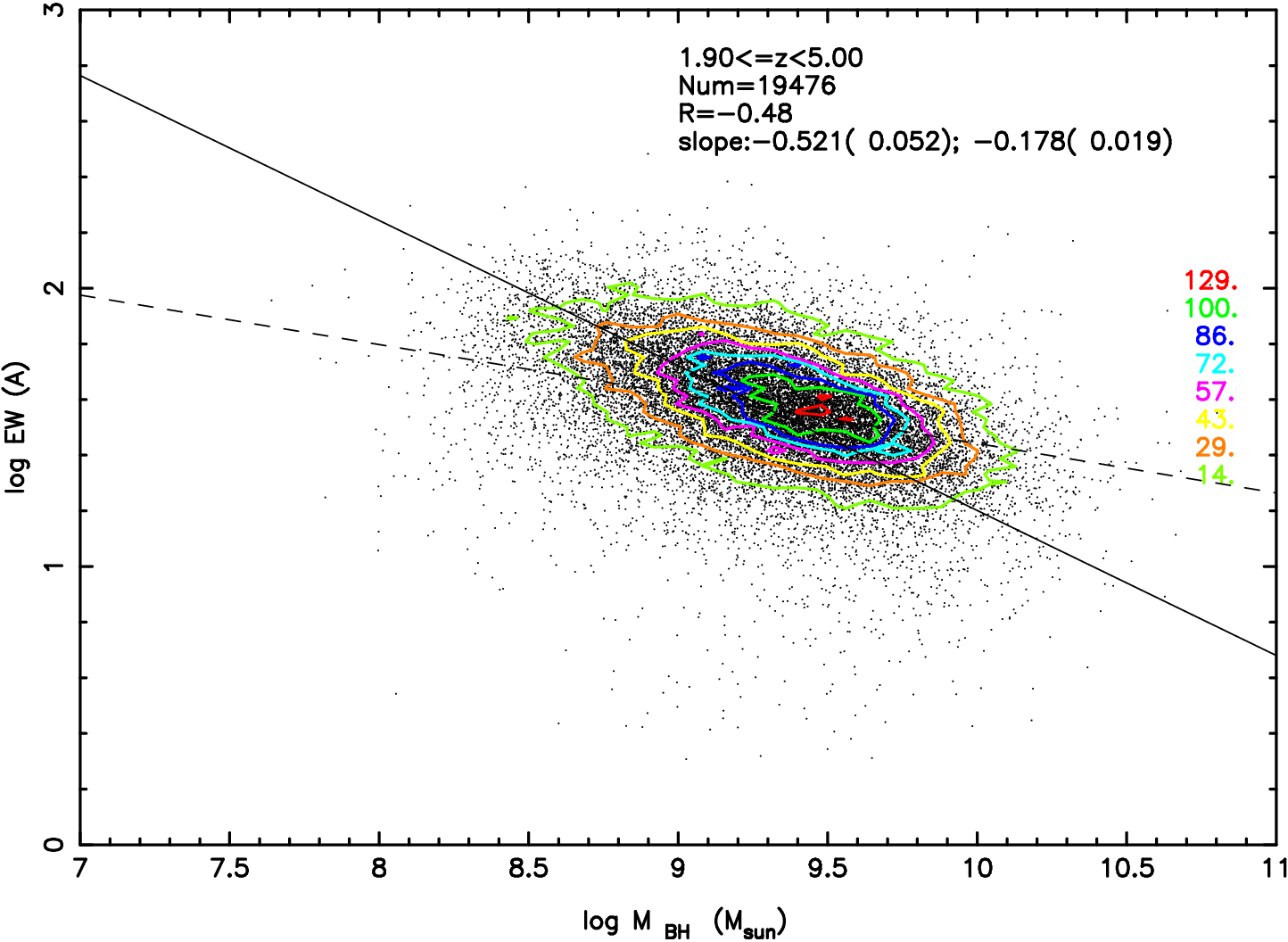} 
\includegraphics[width=6cm,angle=-90]{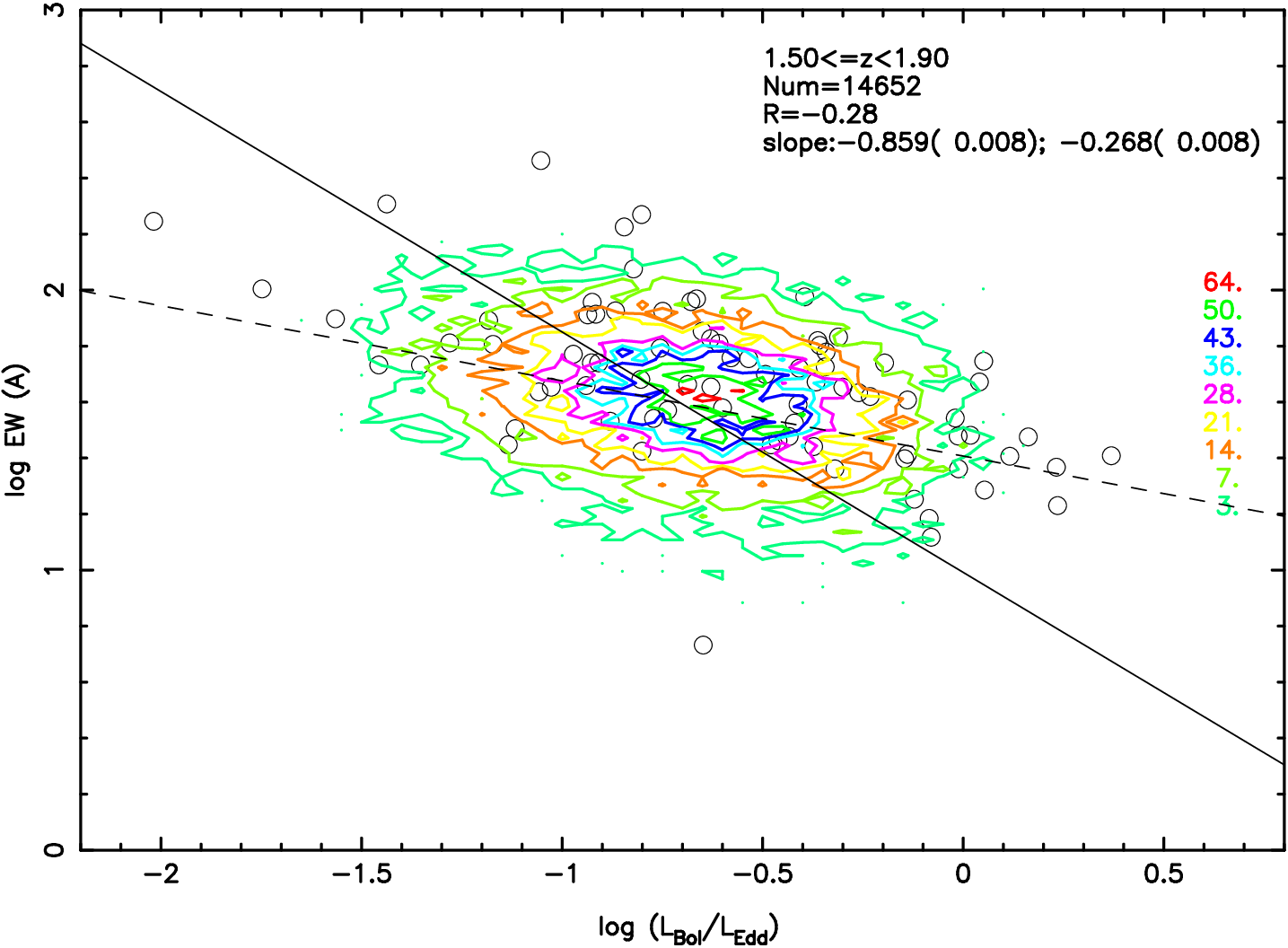} 
\includegraphics[width=6cm,angle=-90]{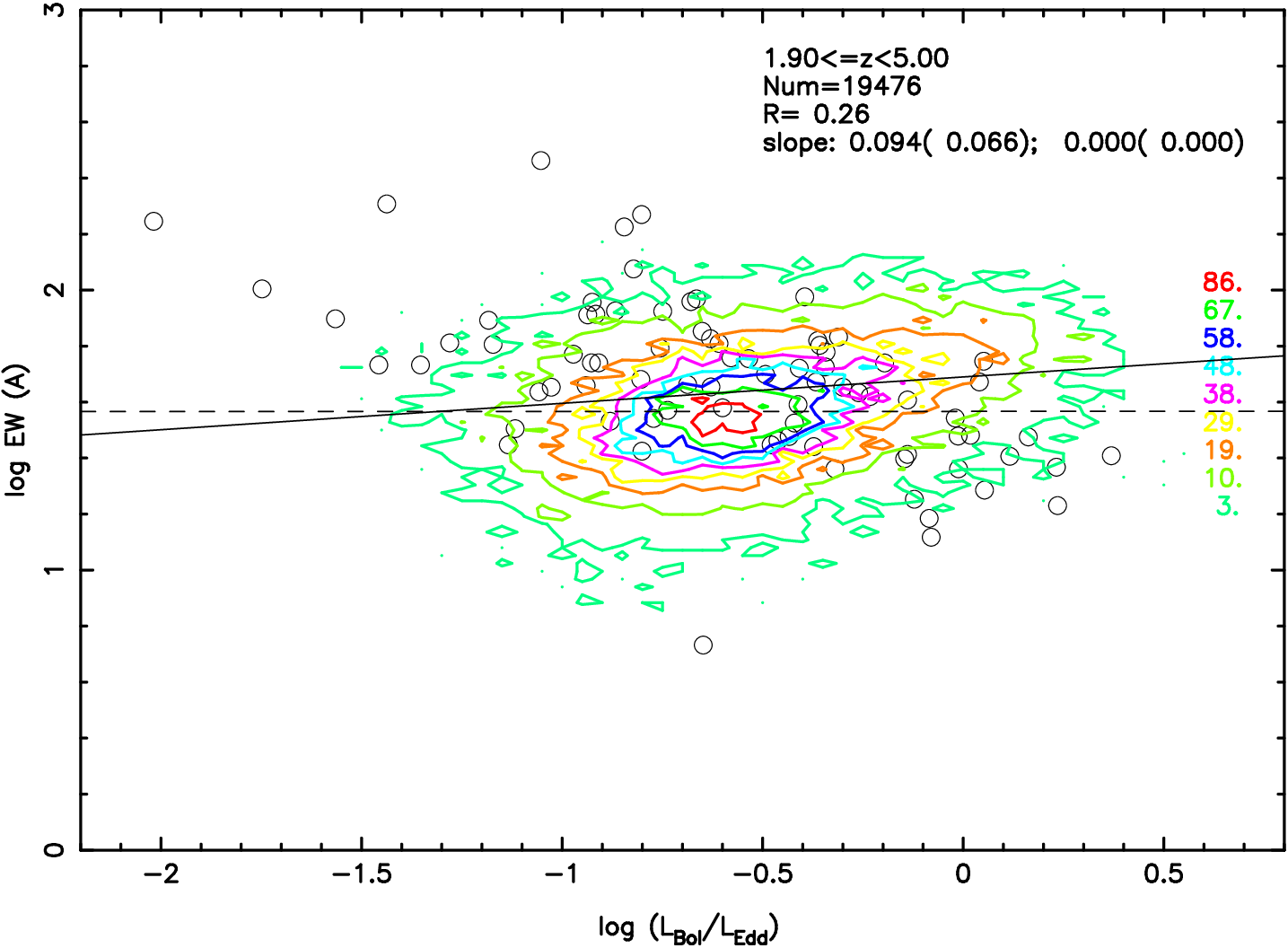} 
\caption{log-log correlation between the rest-frame \civ\ EW and the
\mbh (top), \leddR (bottom) for two redshift bins, $1.50 \leq z <
1.9$, $z\geq 1.9$. In the right of four panels, the numbers for
colorful contour levels are shown. The solid line is the linear fit
from the BCES bisector and the bootstrap simulation. The dash line
is BCES with EW(\civ) as the dependent variable and \mbh/\leddR as
the independent variable. In the right top corner of four panels,
the redshift coverage, the number of the sample, the spearman's rank
correlation coefficient, the slopes and their errors with
BCES($Y|X$) and BCES bisector are listed. Black circles in two
bottom panels are data from Baskin \& Laor (2004).}
\end{center}
\end{figure*}

The underlying physical origin for the Baldwin effect is still an
open question. The fundamental process in QSOs is the accretion
around the SMBHs. Using the data from Shen et al. (2011), we also
investigate the relation between the rest-frame \civ\ EW and \mbh or
\leddR (Figure 5). For the total sample, we find a medium strong
correlation between the rest-frame \civ\ EW and \mbh with $R=-0.27$,
and the null hypothesis is less then $10^{-4}$. However, the
correlation between the \civ\ EW and \leddR is very weak, $R=0.01$
(Figure 5). Earlier studies on individual sources suggest that for
the same object, an intrinsic Baldwin effect does exist (albeit with
a different slope), therefore, luminosity does drive the Baldwin
effect (for the same object, it is equivalent to say that \leddR
drives the Baldwin effect). We find that the very weak correlation
between \civ\ EW and \leddR for the total SDSS DR7 sample is due to
the mixed use of \mgii-based and \civ-based \mbh estimates.

For QSOs with $z\geq 1.9$, we use the mass from the \civ\ line,
while the mass from \mgii\ line for QSOs with $1.5\leq z <1.9$ (Shen
et al. 2011). So, we divide the total our sample into two samples,
i.e., $1.5\leq z <1.9$ and $z\geq 1.9$. For QSOs with $1.5\leq z
<1.9$, We find a very weak correlation of $R=0.07$ between the
rest-frame \civ\ EW and mass from the \mgii\ line. For the physical
drive of the Baldwin effect, \lcon and \mbh are two essentially
independent quantities. To understand which one is a better physical
drive on the Baldwin effect, one needs to look at them separately,
i.e., at fixed \lcon or fixed \mbh. Using the whole sample with
mixed \mbh and \lcon makes it difficult to isolate the two. For
\lcon between $10^{45.7}$ and $10^{46.5}$ \ergs, we divide the total
SDSS sample into four luminosity bins ($\Delta log L_{\rm
1350}=0.2$), and find that this correlation between \civ EW and \mbh
is stronger ($R\sim 0.18$) than that for the total SDSS sample
($R\sim 0.07$). For QSOs with $z\geq 1.9$, we find a strong
correlation of $R=-0.48$ between the \civ\ EW and \civ-based \mbh
(Table 2; Figure 6). It is consistent with our previous result where
we used the \civ\ FWHM from the standard pipeline of SDSS to
calculate the single-epoch \mbh (Vestergaard \& Peterson 2006; Xu et
al. 2008). In section 3.4, the bias in \civ-based \mbh will be
discussed.

For QSOs with $1.5\leq z <1.9$, we find a modest correlation between
\civ\ EW and \leddR ($R=-0.28$). Because \mgii-based \mbh is
proportional to $v^2 \times R_{\rm BLR}$, and $R_{\rm BLR} \propto
\lcon^{0.62}$ (Shen et al. 2011), \leddR is proportional to
$\lcon^{0.38}$ neglecting the effect of \mgii\ FWHM. We find that
$\rm {EW(C IV)} \propto (\leddR)^{-0.27}$ for QSOs with $1.5\leq z
<1.9$ by BCES($Y|X$). Using this correlation, we can derive the
expected Baldwin effect of ${\rm EW(C IV)} \propto \lcon^{-0.10}$,
however, the slope of Baldwin effect (Table 1) is -0.24. It suggests
that the relation between \civ\ EW and \leddR is not completely from
the relation between \civ\ EW and \lcon. We use partial Kendalls
$\tau$ to investigate the role of \lcon in the relation between
\civ\ EW and \leddR for the total sample. We find that the relation
between \civ\ EW and \leddR is weakened when \lcon is held fixed.
The partial Kendalls $\tau$ is $-0.135$ for QSOs with $1.5\leq z
<1.9$, and 0.097 for QSOs with $1.9\leq z <5.0$, while their
spearman's rank correlation coefficients are -0.28 and 0.26 (Table
2). It is possible that \leddR is not very important in controlling
the Baldwin effect, or that the large uncertainty of individual \mbh
estimates is diluting any inherent correlation.

In Figure 6, we also show the location of low-z PG QSOs from Baskin
\& Laor (2004), where the \mbh was calculated from the \hb FWHM.
These PG QSOs, which are denoted by black circles in Figure 6, are
consistent well with our sample of $1.5\leq z <1.9$ QSOs where
\mgii-based \mbh is used. However, these PG QSOs don't follow the
relation found for $1.9\leq z <5.0$ QSOs where \civ-based \mbh is
used. With respect to \hb-based \leddR, it was found that the
relation between EW(\civ) and \leddR is weaker when \leddR is based
on the \civ\ FWHM (Baskin \& Laor 2005). Considering the strong
correlation between \civ\ EW and \leddR for low-z PG QSOs found by
Baskin \& Laor (2004), our result also shows a strong bias of
\civ-based \mbh, with respect to \mgii-based \mbh.

\subsection{The \civ-based \mbh correction from the \civ\ blueshift relative to \mgii}

\begin{figure*}
\begin{center}
\includegraphics[width=6cm,angle=-90]{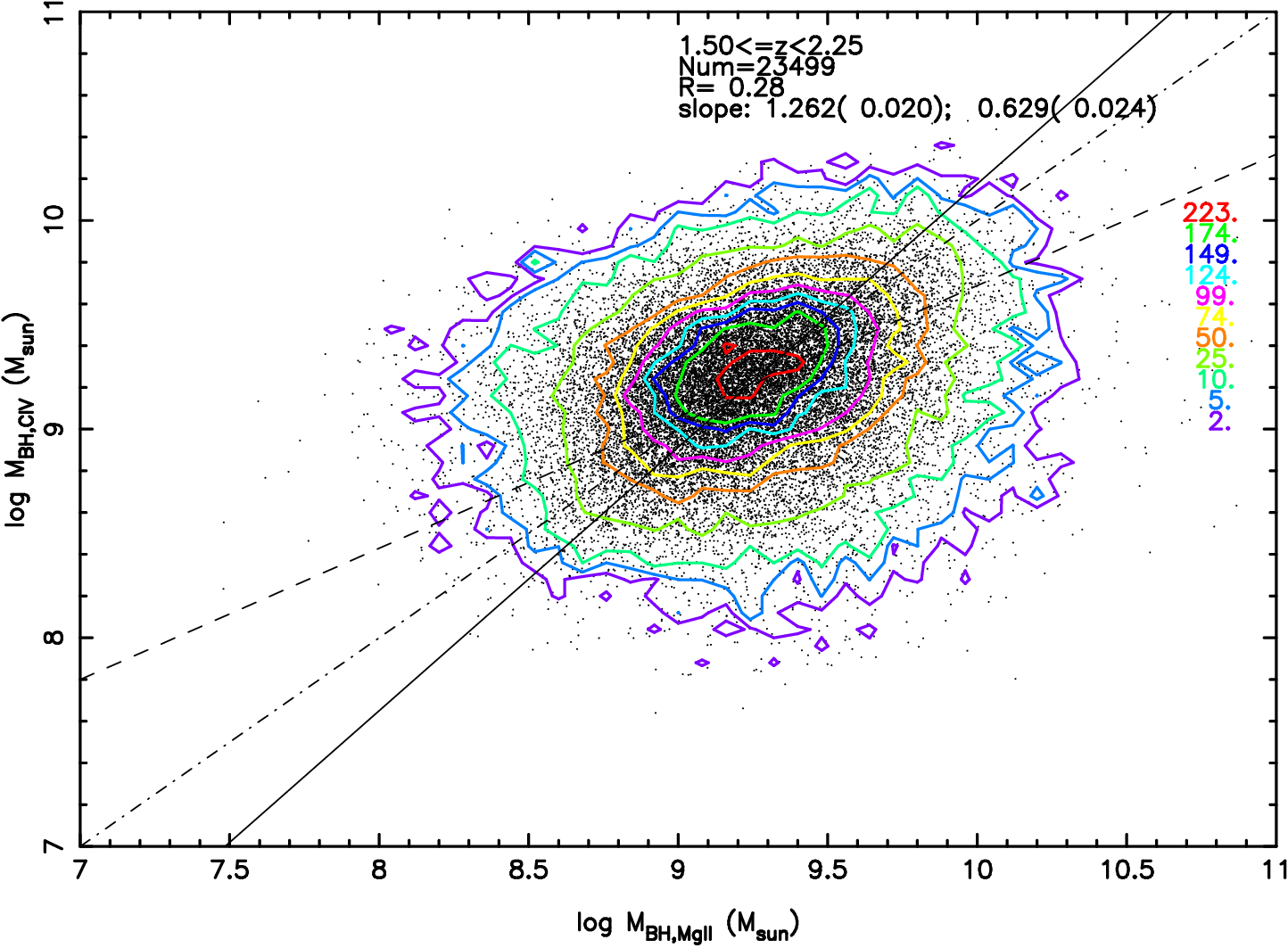}
\includegraphics[width=6cm,angle=-90]{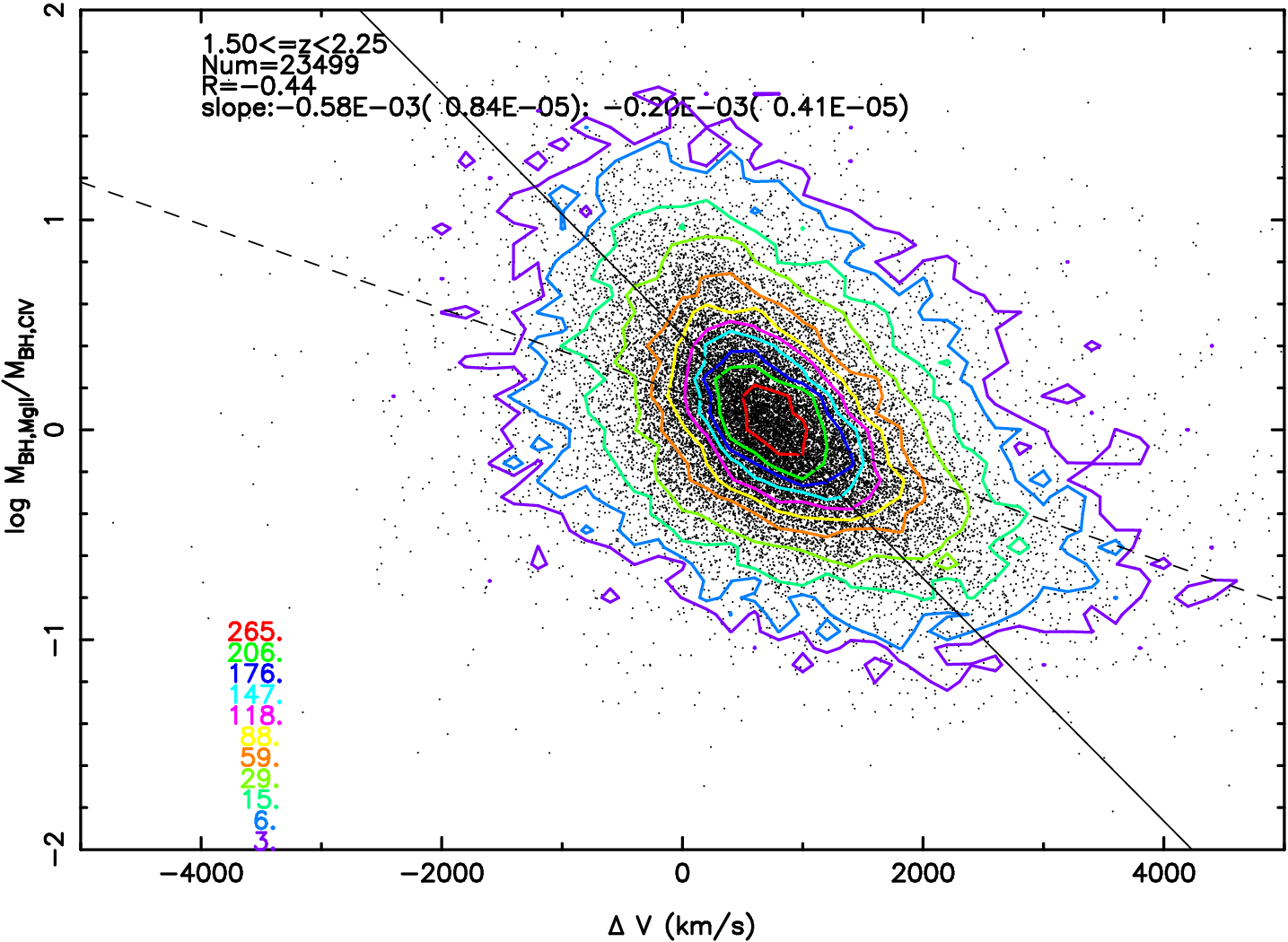}
\includegraphics[width=6cm,angle=-90]{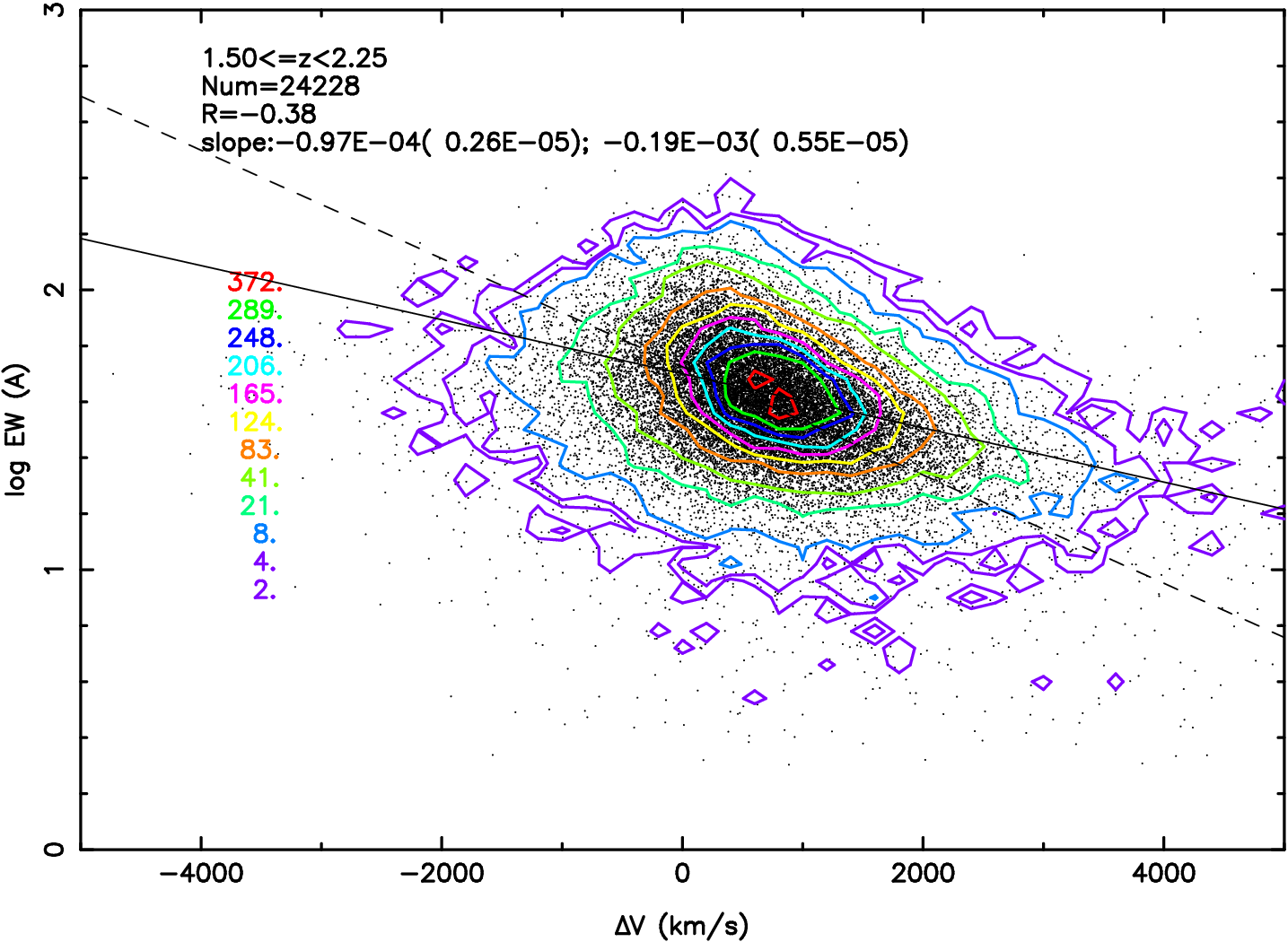}
\includegraphics[width=6cm,angle=-90]{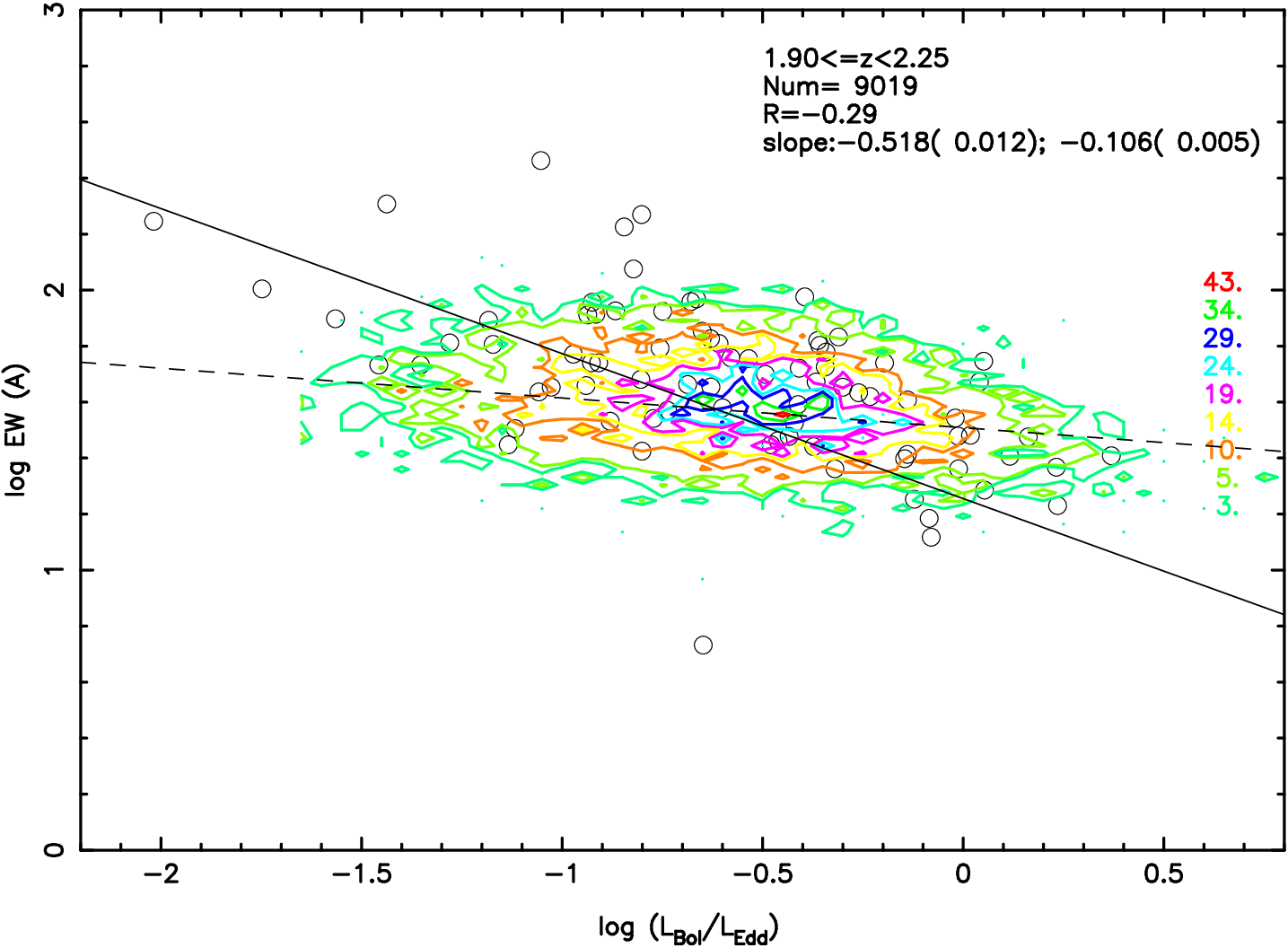}
\caption{Top left: \civ-based \mbh versus \mgii-based \mbh. Top
right: The ratio of \mgii-based \mbh to \civ-based \mbh versus the
\civ\ blueshift relative to \mgii. Bottom left: \civ\ EW versus the
\civ\ blueshift relative to \mgii. Bottom right: \civ\ EW versus the
corrected \civ-based \leddR (for clarity, the data points are not
shown). The numbers for colorful contour levels are shown in
different panels. The solid line is the linear fit from the BCES
bisector and the bootstrap simulation. The dash line is BCES with Y
as the dependent variable and X as the independent variable. In the
top corner of four panels, the redshift coverage, the number of the
sample, the spearman's rank correlation coefficient, the slopes and
their errors with BCES($Y|X$) and BCES bisector are listed. Black
circles in the bottom right panel are data from Baskin \& Laor
(2004).}
\end{center}
\end{figure*}

However, we should keep in mind that the underlying physical drive
analysis depends on the \mbh calculation. The bias of \mbh
calculated from \mgii, \civ\ are investigated (e.g., Rafiee \& Hall
2011; Shen et al. 2012). It is found that the \civ-based \mbh is
biased to the possible non-virialized component in \civ\ (Shen et
al., 2008; Richard et al. 2011; Shen et al. 2011; Shen et al. 2012).
The \mgii-based \mbh can be calibrated to yield consistent virial
mass estimates with those based on the \ha/\hb, while the \civ-based
\mbh is poorly correlated with the \hb-based \mbh or \ha-based \mbh
(Shen et al. 2012). It was found that the \civ\ blueshift relative
to \hb can be used as a \civ\ FWHM correction (shen et al. 2012).
For 24228 QSOs with $1.5\le z \le 2.25$, we find a strong
correlation between \civ-based \mbh and the \civ\ blueshift relative
to \mgii\ ($\Delta V=v_{\rm off,CIV}-v_{\rm off,MgII}$), $R=-0.41$.
We also find a strong correlation between \civ\ EW and the \civ\
blueshift relative to \mgii, $R=-0.38$ (bottom left panel in Figure
7; $R=-0.11$ when only using the \civ\ blueshift). This strong
correlation between \civ\ EW and the \civ\ blueshift relative to
\mgii\ is consistent with the disk+wind model (Richards et al.
2011). It is possible that QSOs with larger \civ\ blueshift have
softer SEDs and have a larger radiation line driving contribution to
their winds to dominate over the disk.

The \civ\ blueshift relative to \mgii\ would lead to the apparent
correlation between \civ\ EW and \civ-based \mbh. We can use the
\civ\ blueshift relative to \mgii\ to do the \civ-based \mbh
correction. Shen et al. (2011) gave the \mgii-based \mbh and
\civ-based \mbh for QSOs with $1.5\le z \le 2.25$ (their Table 1).
Their spearman's rank correlation coefficient is R = 0.28 and the
null hypothesis is less then $10^{-4}$ (top left panel in Figure 7).
We use the \civ\ blueshift relative to \mgii\ to do the correction
of \civ-based \mbh. Using BCES regression analyzes, we find a strong
correlation between the \mbh difference ($\log M_{\rm BH,MgII}- \log
M_{\rm BH,MgII}$) and the \civ\ blueshift relative to \mgii, R=-0.44
(top right panel in Figure 7):
\begin{eqnarray}
\log M_{\rm BH,MgII}-\log M_{\rm BH,CIV}=a+b(v_{\rm off,CIV}-v_{\rm off,MgII})
\end{eqnarray}
Where $a=(0.174\pm 0.004)$, $b=(-2.01\pm 0.04)\times10^{-4}$ by
BCES($Y|X$), and $a=(0.449\pm 0.007)$, $b=(-5.78\pm
0.08)\times10^{-4}$ by BCES bisector.


Considering the correction of \civ-based \mbh from above formula,
the spearman's rank correlation coefficient for the relation of
\mgii-based \mbh and the corrected \civ-based \mbh changes from 0.28
to 0.38, 0.36, respectively. Using above formula by BCES($Y|X$) and
BCES bisector to do the \civ-based \mbh correction, we find that,
for the relation between \civ\ EW and \leddR for 9019 QSOs with
$1.9\le z \le 2.25$, R changes from 0.22 to -0.03, -0.29,
respectively (bottom right panel). And for the relation between
\civ\ EW and the corrected \civ-based \mbh, R changes from -0.45 to
-0.24, 0.1, respectively. They are consistent with the results for 14652
QSOs with $1.5\le z \le 1.9$ where \mgii-based \mbh is used.

\subsection{The origin of the Baldwin effect}

Considering \mgii-based \mbh (QSOs with $z<1.9$ QSOs), we find that
the correlation coefficient for the relation between \civ\ EW and
\lcon ($R=-0.34$) is larger than the other two relations (i.e. \civ\
EW versus \mbh, \civ\ EW versus \leddR; $R=0.07, -0.28$). And the
correlation coefficient of \civ\ EW with \leddR ($R=-0.28$) is
larger than that of \civ\ EW with \mbh ($R=0.07$). It is consistent
with the result by Baskin \& Laor (2004). The stronger correlation
between \civ\ EW and \mbh in Xu et al. (2008) is due to the usage of
\civ-based \mbh. Therefore, by the larger sample, we find that \civ\
EW is primarily controlled by the \lcon, with respect to \leddR and
\mbh. The modest correlation between \civ\ EW and the \leddR (larger
than that between \civ\ EW and \mbh) implied that the \leddR seems
to be a better underlying physical parameter than \mbh. The strong
correlation between the \civ\ EW and \civ\ blueshift relative to
\mgii\ is in favor of the disk+wind model (e.g. Richards et al.
2011). The QSOs with larger \civ\ blueshift have softer SEDs and
have a larger radiation line driving contribution to their winds to
dominate over the disk.


\section{conclusions}
With respect to our previous analysis for SDSS DR5 QSOs, we use the
SDSS DR7 QSOs catalog from Shen et al. (2011) to reinvestigate the
Baldwin effect, its slope evolution, the underlying drive. The main
conclusions can be summarized as follows: (1) The Baldwin effect
exists in the large sample of 35019 QSOs with reliable spectral
analysis. The correlation coefficient is $R=-0.33$, which is almost
the same for QSOs in different redshifts, up to $z\sim 5$. (2) For
the total sample, the slope is -0.238 by the BCES (\civ\ EW depends
on \lcon), -0.787 by the BCES bisector. (3) For 11 redshift bins,
there is an increasing of the Baldwin effect slope from $z\sim 1.5$
to $z\sim2.0$. From $z\sim2.0$ to $z\sim5.0$, the slope change is
not clear considering their uncertainties or the larger redshift
bin. (4) By 34128 QSOs with \mbh and \leddR carefully derived from
spectral decomposition by Shen et al. (2011), we find that there is
a strong correlation between the rest-frame \civ\ EW and \civ-based
\mbh for $1.9\le z < 5.0$($R=-0.48$). However, the relation between
\civ\ EW and \mgii-based \mbh for $1.5\le z <1.9$ is very weak
($R=0.07$). We think it is due to the bias of the \civ-based \mbh,
with respect to the \mgii-based \mbh. Using the correction of
\civ-based \mbh from the \civ\ blueshift relative to \mgii, we find
the correlation of \civ\ EW with corrected \civ-based \mbh becomes
weak (from $R=-0.45$ to $R=0.1$). Considering the \mgii-based \mbh,
a medium strong correlation is found between \civ\ EW and \leddR
($R=-0.28$), which implies that the \leddR seems to be a better
underlying physical parameter than \mbh.

\section{ACKNOWLEDGMENTS}
We thank the discussion during the LAMOST science meeting at April,
2012. We thank an anonymous referee for suggestions that led to
improvements in this paper. This work has been supported by the
National Science Foundations of China (No. 11173016; 11233003).

\end{document}